\documentclass[a4paper,10pt]{article}
\usepackage{graphicx,amsmath,natbib}
\newcommand{\veczan}[1]{\bf {#1}}
\def\htrg{NGC\,1265~}
\def\curvature{R}
\def\circleradius{R_c}
\def\h0units{\mathrm{km\,s^{-1}\,Mpc^{-1}}}

\def\aap{A\&A\,  }%% Astronomy and Astrophysics
\def\aj{AJ  }%% The Astronomical Journal
\def\apj{ApJ\,  }%% Astrophysical Journal
\def\apjl{ApJ\,  }%% Astrophysical Journal, Letters
\def\apss{Astrophysics and Space Science  }%Astrophysics and Space Science
\def\mnras{MNRAS\,  }%% Monthly Notices of the RAS

\title
{
Classical and relativistic
conservation of momentum flux
in radio-galaxies
}
\author{L. Zaninetti   \\Dipartimento di Fisica \\Via Pietro Giuria 1    \\10125, Turin, Italy    \\
\footnote{zaninetti@ph.unito.it}\hspace{0.08cm}
{Corresponding author: zaninetti@ph.unito.it}}
\begin{document}
\maketitle
\begin{abstract}
A new set  of laws of motion for turbulent jets
propagating in
an intergalactic medium characterized by a
decreasing density is derived
by applying
conservation of momentum flux
both in the classical  and relativistic
framework.
Two characteristic features of radio-galaxies,
such as oscillations and
curvature, are modeled by a classical helicoidal jet.
A third feature of a radio-galaxy, the appearance of knots,
is explained
as an effect due to the theory of images.
\end{abstract}
{
\bf{Keywords:}
}
Galaxies:jets
radio continuum : galaxies

\section{Introduction}

The study of extra-galactic jets started with the observations
of NGC 4486 (M87) where
`a curious straight ray lies in a sharp
gap in the nebulosity \ldots',
see \cite{Curtis1918}.
More recently
the analysis  of the radio images  of extra-galactic jets suggests
some interesting problems of physics that should be solved.
One such problem  is that of the decrease in the velocity
as a function of the distance from the parent nucleus,
see \cite{lb2002,laing2002,LaingBridle2004,Laing_2006,Laing2014}.
Various mechanisms have been suggested for the jet's deceleration,
we  report some of them:
an  interaction between a  relativistic jet and the thermal
radiation  from an Active Galactic Nuclei (AGN), see \cite{Melia1989};
the incorporation  of mass
in two-dimensional relativistic jets, see \cite{Bowman1996};
continuous injection of plasma  at the base of the jet
and  dissipation at some distance from the central core,
see \cite{Wang2004};
a rotation-induced Rayleigh--Taylor type instability,
see \cite{Meliani2009}; and
the loading of stellar mass produced in elliptical galaxies,
see \cite{Perucho2014}.
Other authors, such as \cite{Hardcastle2004}, quote
a terminal velocity of $0.3\,c$  and suggest that the
jet's velocity is constant.
A second  problem is that of the physical mechanism
which bends the jets originating in the the head--tail radio-galaxies,
such as
\htrg , see \cite{Owen1978},  or 1159 + 583, see \cite{Burns1979},
or 1638 + 538, see \cite{Burns1980}.
The suggested physical mechanisms for bending are:
trajectories in the independent  blob model, see \cite{jaffe};
an adiabatic model in which the bending is produced
by the ram pressure of a central galaxy
in uniform motion, see \cite{O'Donoghue1993};
the ram pressure in cluster mergers, see \cite{Sakelliou2000}.
The extragalactic  radio-jets are
characterized by a radio-luminosity which is  expressed in watts and
also the flux of kinetic energy is expressed in
watts.
The conversion of the flux of kinetic energy in radio-luminosity
through a turbulent cascade has became an  active
field of research, see \cite{Bicknell1982}.
%In the case of low values of the constant of conversion
%it can be conjectured that the physics  of the jet's propagation
%is independent of the radio-luminosity.
%This conjecture allows the exploration of turbulent jets
%in which, by definition, the matter's  density   is the  same
%as that of the surrounding medium.
%Other approaches assume that  the
% relativistic jets are very dilute,
%        even if mass is loaded and decelerated,
%        compared with the ambient medium, see for example  \cite{lb2002}.

We now  pose
the following questions.
\begin {itemize}
\item
Can the physics of  turbulent jets, which  are observed
in  the laboratory,   be applied  to extra-galactic jets?
\item
Is it possible to generalize the physics of turbulent jets
to a medium with a varying  density?
\item
Can we extend to the relativistic regime the physics of
 turbulent jets?
\item
Can we  explain the curved trajectories of the
extra-galactic jets such as \htrg ?
\item
Is it possible to build the image of a curved  turbulent jet
which  is emitting synchrotron radiation?
\end{itemize}
In order to answer these questions,
we derive the  differential equations
which model the classical and relativistic  momentum conservation
for a  jet in  the presence of three types of medium, see Sections
\ref{secclassic} and
\ref{secrelativistic}.
A model for the composition of a decreasing jet  velocity along the $x$-axis with constant velocity of the hosting galaxy along the $y$-axis
is derived, see
Section \ref{secbented}.
The presence of oscillations is modeled by a helicoidal jet
and a bent helicoidal jet, see Section \ref{sechelicoidal}.
Section \ref{sectionimage} updates an algorithm which allows
building the radio-image of a bent helicoidal jet.

\section{Classical turbulent jets}
\label{secclassic}

\subsection{Luminosity conversion}

The total  power, $Q$, released  in a turbulent
cascade of the Kolmogorov type is, see~\cite{pelletier},
\begin{equation}
Q \approx \gamma_{KH} \rho  s_T^2 \quad ,
\end{equation}
 where
$\gamma_{KH} =  \gamma_{ad}  \frac{s_T}{r}$
is the growth rate of K--H instabilities, $\rho$
is the density of the matter, r is the local radius of the
jet and $s_T$ is the velocity of sound,
see  \cite{Zaninetti2007_b}.
The total maximum luminosity, $L_t$, which  can be
 obtained for
the jet in a given region of radius $r$ and length $r$ is
\begin{equation}
L_t = \pi r^2 r  Q = \gamma_{ad} \rho s_T^3 \pi r^2
\quad  .
\end{equation}
The mechanical luminosity  of  the jet  is
\begin{equation}
L_m  = \frac{1}{2}  \rho v^3 \pi r^2
\quad ,
\label {mechanical}
\end{equation}
and therefore the efficiency
of the conversion, $\chi_T$,
of the total available energy  into
turbulence is
\begin{equation}
\chi_T  = \frac {L_t}{L_m} = \gamma_{ad} \frac{2}{\pi} \frac
{1}{M^3} \approx \frac {1}{M^3} \quad ,
\end{equation}
where $M$ is the Mach number.
It is clear that in a hot jet characterized
by high values of $M$
the fraction of the total available energy released firstly in the
turbulence and after in non-thermal particles is a small fraction
of the bulk flow energy.
The assumption made in
the following,  in which
the bulk flow motion is treated independently from the non-thermal
emission, is now justified.
A similar approach can be found  in \cite{Bicknell1982,Bicknell1984}.

\subsection{The parameters for a turbulent jet}

The physics of a turbulent jet  can be divided into
the simple model and the complex model.
The simple model is characterized  by  an opening angle
$\alpha$, and the matter's density $\rho$ is the same
inside and outside the jet, see Section 35 in \cite{landau}.
The complex model is characterized  by the turbulent viscosity,
$ \nu_T$  and,
as an example,   an opening angle
that is a function of the turbulent viscosity
can be derived
starting from Eq. (5.104) in \cite{Pope2000}
\begin{equation}
\alpha =
2\,\arctan \left( 8\, \left( \sqrt {2}-1 \right) \nu_{{T}} \right)
\quad .
\end{equation}
In the complex model the matter's density $\rho$
is included in   $ \nu_T$.
In both models, the temperature and the pressure are absent
and we can speak of cold jets.

\subsection{Momentum conservation}

The conservation of the momentum flux in a
`turbulent  jet'  as outlined in  \cite {landau}
requires the perpendicular section to the motion along the
Cartesian $x$-axis, $A$
\begin {equation}
A(r)=\pi~r^2
\quad
\end{equation}
where $r$ is the radius of the jet.
The
section  $A$ at  position $x_0$  is
\begin {equation}
A(x_0)=\pi ( x_0   \tan ( \frac{\alpha}{2}))^2
\quad  ,
\end{equation}
where   $\alpha$  is the opening angle and
$x_0$ is the initial position on the $x$-axis.
At position $x$ we have
\begin {equation}
A(x)=\pi ( x   \tan ( \frac{\alpha}{2}))^2
\quad .
\end{equation}
The conservation  of momentum flux states that
\begin{equation}
\rho(x_0)  v_0^2  A(x_0)  =
\rho(x)    v(x)^2 A(x)
\quad  ,
\label{conservazione}
\end {equation}
where $v(x)$ is the velocity at  position $x$ and
$v_0(x_0)$   is the velocity at  position $x_0$.

The selected physical units are
pc for length  and  yr for time;
with these units, the initial velocity $v_{{0}}$
is  expressed in $\mathrm{pc \,yr^{-1}}$,
 1 yr = 365.25 days.
When the initial velocity is expressed in
km\,s$^{-1}$, the multiplicative factor $1.02\times10^{-6}$
should be applied in order to have the velocity expressed in
$\mathrm{pc \,yr^{-1}}$.
The tests are performed on a typical distance
of 15 kpc relative  to the jets in 3C\,31,
see Figure 2 in \cite{laing2002}.

\subsection{Constant  density}

\label{classicalconstant}

In the case of a constant density of the intergalactic medium (IGM)
along the $x$-direction,
the  law of  conservation of the
momentum flux, as given by Eq. (\ref{conservazione}),
can be written as the differential equation
\begin{equation}
 \left( {\frac {\rm d}{{\rm d}t}}x \left( t \right)  \right) ^{2}
 \left( x \left( t \right)  \right) ^{2}-{v_{{0}}}^{2}{x_{{0}}}^{2}=0
\quad .
\end{equation}
The analytical  solution of the previous differential
equation can be found by imposing $x=x_0$ at t=0,
\begin{equation}
x(t) =
\sqrt {2\,tv_{{0}}x_{{0}}+{x_{{0}}}^{2}}
\quad .
\label{xtconstant}
\end{equation}
The asymptotic approximation, see \cite{NIST2010},
is
\begin{equation}
x(t) \sim
\sqrt {2}\sqrt {v_{{0}}x_{{0}}}\sqrt {t}
\quad .
\end{equation}
The velocity is
\begin{equation}
v(t) =
{\frac {v_{{0}}x_{{0}}}{\sqrt {2\,tv_{{0}}x_{{0}}+{x_{{0}}}^{2}}}}
\quad  ,
\end{equation}
and its asymptotic approximation
\begin{equation}
v(t) \sim
\frac
{
\sqrt {2}\sqrt {v_{{0}}}\sqrt {x_{{0}}}
}
{
2\,\sqrt {t}
}
\quad  .
\end{equation}
The transit time, $t_{tr}$, necessary to travel a distance
$x_{max}$ can be derived from Eq. (\ref{xtconstant})
\begin{equation}
t_{tr} = \frac
{
-{x_{{0}}}^{2}+{x_{{\max}}}^{2}
}
{
2\,v_{{0}}x_{{0}}
}
\quad .
\end{equation}
As a numerical  example, inserting
$x_0$=100 pc, $x_{max}=15\,10^3\,pc=15$\ kpc, which is the reference value,
and  a classical initial velocity of
$v_0=$10000 km/s (
$v_0=0.0102\,$pc/yr),
we obtain  $t_{tr} =1.1\,10^8\,$yr.

\subsection{An hyperbolic  profile of the density}

The density  is  assumed to decrease as
\begin{equation}
\rho = \rho_0  (\frac{x_0}{x})
\quad ,
\label{profhyperbolic}
\end{equation}
where  $\rho_0=0$ is the density at  $x=x_0$.
The differential equation that models the
momentum flux is
\begin{equation}
x \left( t \right)  \left( {\frac {\rm d}{{\rm d}t}}x \left( t
 \right)  \right) ^{2}-{v_{{0}}}^{2}x_{{0}}=0
\quad ,
\end{equation}
and its analytical solution
is
\begin{equation}
x(t) =
\frac{1}{4}\,{\frac { \left( 12\,tv_{{0}}{x_{{0}}}^{2}+8\,{x_{{0}}}^{3}
 \right) ^{2/3}}{x_{{0}}}}
\quad .
\label{xthyperbolic}
\end{equation}
The asymptotic  approximation
is
\begin{equation}
x(t)  \sim
\frac{1}{4}\,{\frac {{12}^{2/3} \left( v_{{0}}{x_{{0}}}^{2} \right) ^{2/3}{t}^
{2/3}}{x_{{0}}}}
\quad .
\label{xthyperbolicasympt}
\end{equation}
The analytical solution for the velocity
is
\begin{equation}
v(t) =
2\,{\frac {x_{{0}}v_{{0}}}{\sqrt [3]{12\,tv_{{0}}{x_{{0}}}^{2}+8\,{x_{
{0}}}^{3}}}}
\quad ,
\end{equation}
and its asymptotic  approximation is
\begin{equation}
v(t) \sim
\frac
{
{12}^{2/3}{v_{{0}}}^{2/3}\sqrt [3]{x_{{0}}}
}
{
6\,\sqrt [3]{t}
}
\quad  .
\end{equation}

The transit time can be  derived from
Eq. (\ref{xthyperbolic})
\begin{equation}
t_{tr} = \frac
{
-2\,{x_{{0}}}^{3/2}+2\,{x_{{\max}}}^{3/2}
}
{
3\,\sqrt {x_{{0}}}v_{{0}}
}
\quad ,
\end{equation}
and with the same parameters  as in Section
\ref{classicalconstant},
we have  $t_{tr} =1.19\,10^7\,$yr.

\subsection{An inverse power law  profile of the density}

\begin{figure*}
\begin{center}
\includegraphics[width=7cm]{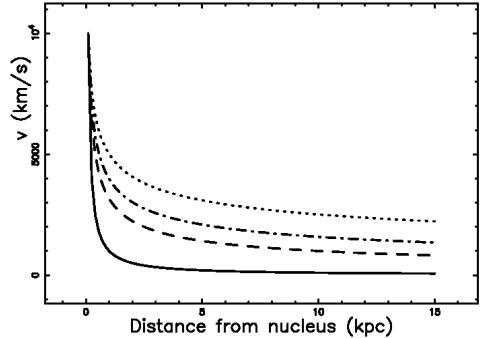}
\end {center}
\caption
{
Classical velocity   as a function
of  the distance from the nucleus  when
$x_0$ =100~pc and $v_0=10000\,$km/s:
$\delta =0$   (full line),
$\delta =1$   (dashes),
$\delta =1.2$ (dot-dash-dot-dash)
and
$\delta =1.4$ (dotted).
}
\label{vxdelta}
    \end{figure*}

The density  is  assumed to decrease as
\begin{equation}
\rho = \rho_0  (\frac{x_0}{x})^{\delta}
\quad ,
\label{profpower}
\end{equation}
where  $\rho_0$ is the density at  $x=x_0$.
The differential equation that models the
momentum flux is
\begin{equation}
\left( {\frac {x_{{0}}}{x}} \right) ^{\delta}{v}^{2}{x}^{2}-{v_{{0}}}
=0
\quad  .
\end{equation}
There is no analytical solution, and we simply express
the velocity as a function of the position, $x$,
\begin{equation}
v(x) =
{\frac {x_{{0}}v_{{0}}}{x}{\frac {1}{\sqrt { \left( {\frac {x_{{0}}}{x
}} \right) ^{\delta}}}}}
\quad ,
\label{velocityhyper}
\end{equation}
see  Figure \ref{vxdelta}.
% figure   vxdelta
% end vxdelta

\section{Relativistic turbulent jets}
\label{secrelativistic}
The conservation of the   momentum flux in special relativity,
 SR,
in  the presence of the velocity $v$ along one direction
states that
\begin{equation}
(w (\frac{v}{c})^2 \frac { 1}{ 1 -\frac {v^2}{c^2} } +p) A(x) = cost
\quad ,
\label{enthalpy}
\end{equation}
where $A(x)$ is the considered area in the direction perpendicular
to the motion,
$c$ is the speed of light,
and  $w$  is  the enthalpy per unit volume
\begin{equation}
w= c^2 \rho  + p
\quad ,
\end{equation}
where $p$ is the pressure,
see \cite{Gourgoulhon2006} or formula A30 in
\cite{deyoung}.
In accordance with the current models of classical turbulent jets,
we  insert $p=0$  and
the   conservation law
for relativistic momentum flux
is
\begin{equation}
(\rho  v^2 \frac { 1}{ 1 -\frac {v^2}{c^2} }) A(x) = cost
\quad .
\end{equation}
Our physical units are
pc for length  and  yr for time, and
in these units, the speed of light is
$c=0.306$  \ pc \ yr$^{-1}$.

\subsection{Constant  density in SR}

The  conservation of the relativistic momentum flux
when the density is constant can be written
as the differential equation
\begin{eqnarray}
{ \left( {\frac {\rm d}{{\rm d}t}}x \left( t \right)
 \right) ^{2}\pi   \left( x \left( t \right)  \right) ^{2} \left(
\tan \left( \frac{\alpha}{2} \right)  \right) ^{2} \left( 1-{\frac { \left( {
\frac {\rm d}{{\rm d}t}}x \left( t \right)  \right) ^{2}}{{c}^{2}}}
 \right) ^{-1}} \nonumber \\
 -{{v_{{0}}}^{2}\pi  {{\it x_0}}^{2} \left(
\tan \left(  \frac{\alpha}{2}  \right)  \right) ^{2} \left( 1-{\frac {{v_{{0}}}^
{2}}{{c}^{2}}} \right) ^{-1}}=0
\label{eqndiffrel}
\quad  .
\end{eqnarray}
An analytical solution of the previous differential equations
at the moment of writing does not
exist but  we can
provide a  power series solution of the form
\begin{equation}
x(t) = a_0 +a_1  t +a_2 t^2 +a_3  t^3 + \dots
\quad ,
\label{rtseries}
\end{equation}
see  \cite{Tenenbaum1963,Ince2012}.
The coefficients $a_n$ up to order 4  are
\begin{eqnarray}
a_0=&  x_{{0}}      \nonumber \\
a_1=&  \beta_{{0}}c     \nonumber  \\
a_2=&  \frac{1}{6} \,{\frac
{3\,{c}^{2}{\beta_{{0}}}^{4}x_{{0}}-3\,{c}^{2}{\beta_{{0}}
}^{2}x_{{0}}}{{x_{{0}}}^{2}}}
    \nonumber \\
a_3=&
\frac{1}{6}\,{\frac {4\,{c}^{3}{\beta_{{0}}}^{7}-7\,{c}^{3}{\beta_{{0}}}^{5}+3
\,{c}^{3}{\beta_{{0}}}^{3}}{{x_{{0}}}^{2}}}
\quad ,
\end{eqnarray}
where $\beta_0 = \frac{v_0}{c}$.
%siamoqui

\begin{figure*}
\begin{center}
\includegraphics[width=7cm]{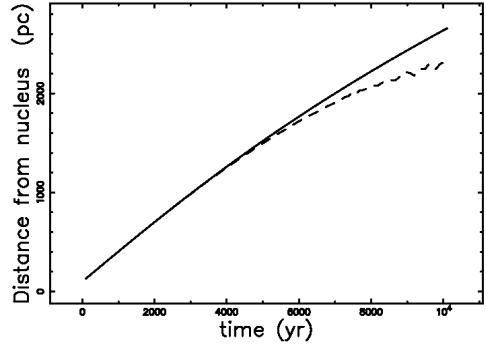}
\end {center}
\caption
{
Non-linear solution  as given
by Eq. (\ref{solrelativisticnl}) (full line)
and series solution
 as given
by Eq. (\ref{rtseries}) (dashed line)
when
$x_0$ =100 pc and $\beta_0$ =0.999.
}
\label{seriesnlrel}
    \end{figure*}

In order to find  a numerical solution of the above
differential  equation
we isolate the velocity from
eq.(\ref{eqndiffrel})
\begin{equation}
v(x;x_0,\beta_0,c) =
{\frac {x_{{0}}\beta_{{0}}{c}^{2}}{\sqrt {-{c}^{2}{x}^{2}{\beta_{{0}}}
^{2}+{c}^{2}{\beta_{{0}}}^{2}{x_{{0}}}^{2}+{c}^{2}{x}^{2}}}}
\,
\end{equation}
and  separate the variables
\begin{equation}
\int_{x_0}^x
{\frac {\sqrt {-{c}^{2}{x}^{2}{\beta_{{0}}}^{2}+{c}^{2}{\beta_{{0}}}^{
2}{x_{{0}}}^{2}+{c}^{2}{x}^{2}}}{x_{{0}}\beta_{{0}}{c}^{2}}}
dx  = \int_0^t  dt
\quad .
\end{equation}
The integral on the left side of the previous equation
has an analytical solution and the
following non-linear equation is obtained
\begin{equation}
\frac{AN}{AD}=t
\quad ,
\label{solrelativisticnl}
\end{equation}
where
\begin{eqnarray}
AN=
{\beta_{{0}}}^{2}{x_{{0}}}^{2}\ln  \left( \sqrt {-{\beta_{{0}}}^{2}+1}
x+\sqrt {-{x}^{2}{\beta_{{0}}}^{2}+{\beta_{{0}}}^{2}{x_{{0}}}^{2}+{x}^
{2}} \right) & ~
\nonumber \\
-{\beta_{{0}}}^{2}{x_{{0}}}^{2}\ln  \left( \sqrt {-{\beta
_{{0}}}^{2}+1}+1 \right) &~
\nonumber \\
 -{\beta_{{0}}}^{2}{x_{{0}}}^{2}\ln  \left( x_
{{0}} \right) +x\sqrt {-{x}^{2}{\beta_{{0}}}^{2}+{\beta_{{0}}}^{2}{x_{
{0}}}^{2}+{x}^{2}}\sqrt {-{\beta_{{0}}}^{2}+1}
&~
\nonumber \\
-{x_{{0}}}^{2}\sqrt {-{
\beta_{{0}}}^{2}+1} &~
\quad ,
\end{eqnarray}
and
\begin{equation}
AD=
2\,\sqrt {-{\beta_{{0}}}^{2}+1}x_{{0}}\beta_{{0}}
\quad .
\end{equation}
Figure \ref{seriesnlrel}  reports  an example of the
above  numerical solution.
%figure seriesnlrel
% end figure seriesnlrel

More details on the case of constant density can be found
in \cite{Zaninetti2009b}.

\subsection{A hyperbolic density profile in SR}

The  conservation of the relativistic momentum flux
in the presence  of an hyperbolic  density profile
as given by Eq. (\ref{profhyperbolic})
is
\begin{eqnarray}
{
{\it x_0}\,x \left( t \right)  \left( {\frac {\rm d}{{\rm d}
t}}x \left( t \right)  \right) ^{2}\pi \, \left( \tan \left( \alpha/2
 \right)  \right) ^{2} \left( 1-{\frac { \left( {\frac {\rm d}{{\rm d}
t}}x \left( t \right)  \right) ^{2}}{{c}^{2}}} \right) ^{-1}}
\nonumber \\
-{
{v_{{0}}}^{2}\pi \,{{\it x_0}}^{2} \left( \tan \left( \alpha/2
 \right)  \right) ^{2} \left( 1-{\frac {{v_{{0}}}^{2}}{{c}^{2}}}
 \right) ^{-1}}=0
\quad  .
\end{eqnarray}
The analytical  solution  of the  previous  differential
equation is
\begin{equation}
x(t) =
\frac
{
-\sqrt [3]{2}\sqrt [3]{x_{{0}}}\sqrt [3]{ \left( 3\,ct{\beta_{{0}}}^{3
}-3\,ct\beta_{{0}}-2\,x_{{0}} \right) ^{2}}+2\,{\beta_{{0}}}^{2}x_{{0}
}
}
{
2\,{\beta_{{0}}}^{2}-2
}
\quad .
\label{xtrelativistic}
\end{equation}
The asymptotic  approximation
is
\begin{eqnarray}
x(t)  \sim
-\frac{1}{2}\,{\frac {\sqrt [3]{2}\sqrt [3]{x_{{0}}}\sqrt [3]{9}\sqrt [3]{{
\beta_{{0}}}^{2}{c}^{2} \left( {\beta_{{0}}}^{2}-1 \right) ^{2}}}{
 \left( {\beta_{{0}}}^{2}-1 \right)  \left( {t}^{-1} \right) ^{2/3}}}+
{\frac {{\beta_{{0}}}^{2}x_{{0}}}{{\beta_{{0}}}^{2}-1}}
\nonumber \\
+ \frac{2}{9} \,{\frac {
\sqrt [3]{2}{x_{{0}}}^{4/3}\sqrt [3]{9}\sqrt [3]{{\beta_{{0}}}^{2}{c}^
{2} \left( {\beta_{{0}}}^{2}-1 \right) ^{2}}\sqrt [3]{{t}^{-1}}}{\beta
_{{0}}c \left( {\beta_{{0}}}^{2}-1 \right) ^{2}}}
\quad  .
\label{xtrelativistichyperbolic}
\end{eqnarray}
The analytical solution when the velocity is expressed as
$\beta=v/c$ is
\begin{equation}
\beta(t) = \frac
{
-\beta_0\, \left( 3\,{\beta_0}^{3}ct-3\,\beta_0\,ct-2\,{\it x_0} \right)
\sqrt [3]{{\it x_0}}\sqrt [3]{2}
}
{
 \left(  \left( 3\,{\beta_0}^{3}ct-3\,\beta_0\,ct-2\,{\it x_0} \right) ^{
2} \right) ^{2/3}
}
\quad ,
\end{equation}
and the asymptotic approximation is
\begin{equation}
\beta(t) = \frac
{
-{\beta_0}^{2}c \left( {\beta_0}^{2}-1 \right) \sqrt [3]{{\it x_0}}\sqrt
[3]{2}\sqrt [3]{9}\sqrt [3]{{t}^{-1}}
}
{
3\, \left( {\beta_0}^{2}{c}^{2} \left( {\beta_0}^{2}-1 \right) ^{2}
 \right) ^{2/3}
}
\quad .
\end{equation}

\subsection{Inverse power law  profile of density in SR }

\begin{figure*}
\begin{center}
\includegraphics[width=7cm]{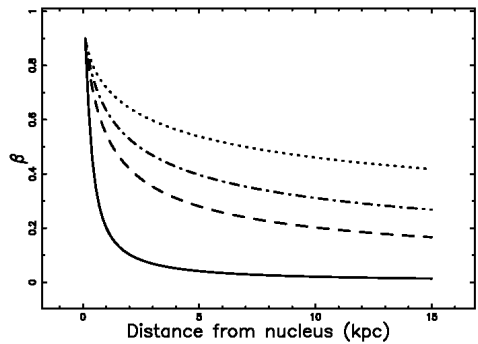}
\end {center}
\caption
{
Relativistic variable $\beta$   as a function
of  the distance from the nucleus  when
$x_0$ =100~pc and $\beta_0$ =0.9:
$\delta =0$   (full line),
$\delta =1$   (dashes),
$\delta =1.2$ (dot-dash-dot-dash)
and
$\delta =1.4$ (dotted).
}
\label{betaxdelta}
    \end{figure*}

The  conservation of the relativistic momentum flux
in the presence  of a
density profile  of the inverse power law type
as given by Eq. (\ref{profpower})
is
\begin{equation}
\frac
{
-{x_{{0}}}^{\delta}{x}^{-\delta+2}{\beta_{{0}}}^{2}{\beta}^{2}+{\beta_
{{0}}}^{2}{\beta}^{2}{x_{{0}}}^{2}+{x_{{0}}}^{\delta}{x}^{-\delta+2}{
\beta}^{2}-{\beta_{{0}}}^{2}{x_{{0}}}^{2}
}
{
\left( {\beta}^{2}-1 \right)  \left( {\beta_{{0}}}^{2}-1 \right)
} =0
\quad .
\end{equation}
This  differential equation does not have an analytical
solution and the expression for $\beta$ as a function of the distance
is
\begin{equation}
\beta(x) = \frac
{
x_{{0}}\beta_{{0}}
}
{
\sqrt {-{\beta_{{0}}}^{2}{x}^{2} \left( {\frac {x_{{0}}}{x}} \right) ^
{\delta}+{\beta_{{0}}}^{2}{x_{{0}}}^{2}+{x}^{2} \left( {\frac {x_{{0}}
}{x}} \right) ^{\delta}}
}
\quad .
\label{betadistance}
\end{equation}
The behavior of $\beta$ as a function of the distance
for different values of $\delta$ can be seen
in Figure \ref{betaxdelta}.
% figure   betaxdelta
% end betaxdelta

The transit time can be  derived from
Eq. (\ref{xtrelativistichyperbolic})
\begin{equation}
t_{tr} = \frac
{
2\,{{\it x_0}}^{2}-2\,\sqrt {{\it x_0}\, \left( {\beta_0}^{2}{\it x_0}-{
\beta_0}^{2}{\it x_{max}}+{\it x_{max}} \right) ^{3}}
}
{
3\,\beta_0\,c \left( {\beta_0}^{2}-1 \right) {\it x_0}
}
\quad  .
\end{equation}
As an example, inserting
$x_0$=100 pc, $x_{max}=15\,10^3\,$pc, and
$\beta_0=0.9$,
we have  $t_{tr} =2\,10^5\,$yr.

\section{Classical curved trajectories}
\label{secbented}

In the presence of a 3D trajectory, ${\veczan{r(t)}}$,
the acceleration is given by
\begin{equation}
{\veczan{a}} = \frac{dv}{dt} {\veczan{T}} +\frac{v^2}{\rho}{\veczan{N}}
\quad ,
\end{equation}
where $v=|\frac{d{\veczan{r(t)}} }{dt}|=|\frac{ds}{dt}$
is the
magnitude of the velocity,
$s$ is the arc-length of the trajectory,
${\veczan{T}}$ and  ${\veczan{N}}$ are the
tangential and normal versors to the trajectory,
and $\curvature$ is the radius of curvature given by
\begin{equation}
\curvature= \left ( (\frac{d^2x}{ds^2})^2 + (\frac{d^2y}{ds^2})^2 +
(\frac{d^2z}{ds^2})^2  \right )^{-1/2}
\quad ,
\label{radiuscurvature}
\end{equation}
see exercise 5.30 in \cite{Spiegel1971}.
A second definition of the radius of curvature
when we have a 2D curve  parametrized as $y(x)$
is
\begin{equation}
\curvature = \frac{ \left ( 1
+ \left( \frac{dy}{dx} \right)^2 \right) ^{\frac{3}{2}} }{
\frac{d^2 y}{dx^2} }
\label{rhogeometrical}
\quad  ,
\end{equation}
see Eq. (12.5) in  \cite{Granville1911}.
In the presence of a trajectory given by a discrete set of points,
the circle of curvature  which has radius   $\circleradius $,
and coordinates of its centre $(x_c,y_c)$,
can be drawn when three points
are given,   see  Appendix \ref{appendice}.
According to theorem 14.1.1 in \cite{Granville1911},
the radius of the circle of curvature and
the radius of curvature are equal.

\subsection{The astronomical data}

\begin{figure*}
\begin{center}
\includegraphics[width=7cm]{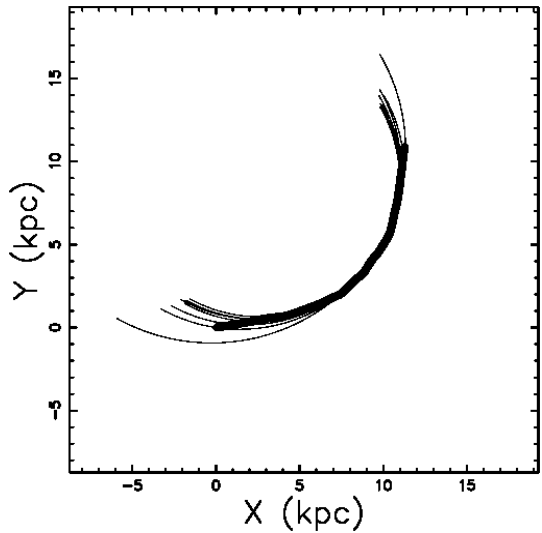}
\end {center}
\caption
{
The right side of  \htrg,
the line with  a large width, represents the real data
as extracted by the author from  Figure 3 in
Owen, Burns, $\&$  Rudnick (1978)
and
8 circles of curvature.
}
\label{ngc1265_circle}
    \end{figure*}

As an application, we analyse  the first part of the right side
of \htrg  where the distances were evaluated
adopting  $H_0 = 50  \h0units $  for the Hubble constant
and $z= 0.0183$ for the redshift
see \cite{Xu1999}.
At the moment of writing, there is a more precise evaluation of
the Hubble constant, which is
$H_0 =(69.6 \pm 0.7 ) \h0units $,
see \cite{Bennett2014},
which, coupled with
$c=299792.458\mathrm{\frac{km}{s}}$  for the velocity of light,
 see \cite{CODATA2012},
 means
\begin{equation}
1^{\prime \prime} = 382.15 pc
\quad .
\label{conversionpppc}
\end{equation}
The trajectory of \htrg was derived as follows
\begin{enumerate}
\item
The data were digitized in $~^{\prime \prime}$ from Figure 3 in
\cite{Owen1978} using WebPlotDigitizer, a
Web based tool to extract data from plots.
\item
The conversion from $~^{\prime \prime}$ to pc was done using
Eq. (\ref{conversionpppc}).
\end{enumerate}
Figure \ref{ngc1265_circle} shows the digitalized 2D trajectory
and 8 circles of curvature computed according to
Eqs (\ref{circleradius}, \ref{circlexc},\ref{circleyc}).
% figure   ngc1265_circle
% end ngc1265_circle

The average radius of the circle of curvature,
$\overline{\circleradius}$, for
the first 20 kpc of the right side
of \htrg is
\begin{equation}
\overline{\circleradius}= (9.22 \pm 1.03) kpc
\quad .
\label{circleradiusng1265}
\end{equation}

\subsection{Curved trajectory  with constant density}

A  ballistic trajectory is determined on
Earth by gravity and aerodynamic
drag.
Here we assume a non-ballistic trajectory for the radio-galaxy
as being due to the composition of
the jet's motion along the $x$-axis
with a motion
at  constant velocity,
$v_g$,
of the parent galaxy along the $y$-axis.
The  peculiar velocity for a galaxy with redshift $z$
can be evaluated using formula (3) in \cite{Freeland2008}.
Other authors replace the velocity  of the parent galaxy
with the velocity of external winds, see \cite{Hardee2003,Perkins2004}.

%We suggest that  the galaxy's velocity in
%the IGM  can be the explanation of
%the relativistic correction for frequencies in
%the $~^7L^+$ relativistic ions   in the storage ring ESR at Darmstadt,
%see \cite{Botermann2014};
% a velocity of our Galaxy of  $\approx 500 \frac{km}{s}$ explains the
%correction, see \cite{Cahill2015}.

\begin{figure*}
\begin{center}
\includegraphics[width=7cm]{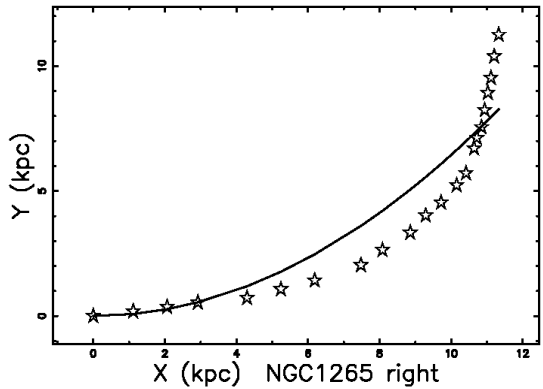}
\end {center}
\caption
{
Real trajectory  of  \htrg (big stars)
and theoretical curved trajectory (full line) in the case of constant density.
The parameters are
$x_0$=100\,pc,
$y_0$=0\,pc,
$v_0=$9620\, km/s
and
$v_{g}=$ 124 km/s.
}
\label{ngc1265_traj}
    \end{figure*}

When the galaxy's  velocity is expressed in
km\,s$^{-1}$, the multiplicative factor $1.02\times10^{-6}$
should be applied in order to have the galaxy's velocity expressed in
$\mathrm{pc \,yr^{-1}}$.
The two equations of motion are
\begin{subequations}
\begin{align}
x(t)  = \sqrt {2\,tv_{{0}}x_{{0}}+{x_{{0}}}^{2}}
\label{xft}     \\
y(t)  = y_0 +v_g t \label{yft}
\quad  ,
\end{align}
\end{subequations}
where $y_0$ is the $y$-position at $t=0$.
In order to evaluate the radius of curvature
in the case of constant  density, we
extract  $t$ from Eq. (\ref{xft})
and we insert  it into Eq. (\ref{yft})
\begin{equation}
y(x) =
\frac{1}{2}\,{\frac {v_{{g}} \left( {x}^{2}-{x_{{0}}}^{2} \right) }{v_{{0}}x_{
{0}}}}+y_{{0}}
\quad .
\label{yxconstant}
\end{equation}
The previous formula allows visualizing
the trajectory
which represents the best fit
for the right side
of \htrg in the case of constant density,
see Figure \ref{ngc1265_traj}.
% figure   ngc1265_traj
% end ngc1265_traj

The parameters $v_0$ and $v_{g}$   are found
by minimizing the value of  $\chi^2$
defined as
\begin{equation}
\chi^2 = \sum_{i=1}^n (y_i -y_{i,th})^2
\quad  ,
\end{equation}
where $y_i$      is the $y$-value of the  $i$th point
in the digitized trajectory
and   $y_{i,th}$ is the theoretical point obtained
by inserting  the $x$ value of the  $i$th point in
Eq. (\ref{yxconstant}).

The radius of curvature in the case of constant density
as  given by
Eq. (\ref{radiuscurvature})
is
\begin{equation}
\curvature=
{\frac { \left( {x}^{2}{v_{{g}}}^{2}+{v_{{0}}}^{2}{x_{{0}}}^{2}
 \right) ^{3/2}}{{v_{{0}}}^{2}{x_{{0}}}^{2}v_{{g}}}}
\quad ,
\end{equation}
as a consequence the  centripetal acceleration is
\begin{equation}
\frac{v^2}{\curvature}{\veczan{N}}
=
{\frac { \left( 2\,tv_{{0}}x_{{0}}+{x_{{0}}}^{2}+ \left( tv_{{g}}+y_{{0
}} \right) ^{2} \right) {v_{{0}}}^{2}{x_{{0}}}^{2}v_{{g}}}{ \left( {v_
{{0}}}^{2}{x_{{0}}}^{2}+{v_{{g}}}^{2} \left( 2\,tv_{{0}}x_{{0}}+{x_{{0
}}}^{2} \right)  \right) ^{3/2}}} {\veczan{N}}
\quad .
\end{equation}

\subsection{Curved trajectory  with hyperbolic density}

\begin{figure*}
\begin{center}
\includegraphics[width=7cm]{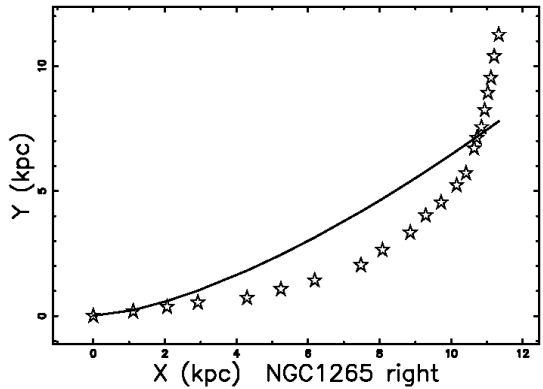}
\end {center}
\caption
{
Real trajectory  of  \htrg (big stars)
and theoretical curved trajectory (full line) in the case
of an hyperbolic  profile of density.
The parameters are
$x_0$=100\, pc,
$y_0$=0  \, pc,
$v_0=$3660 km/s
and
$v_{g}=$54.74 km/s.
}
\label{ngc1265_traj_power1}
    \end{figure*}

In the case of an hyperbolic  density profile
we can isolate the time in Eq. (\ref{xthyperbolic})
and insert it into Eq. (\ref{yft}).
The resulting  trajectory,  which is independent of the time, is
\begin{equation}
y(x) =
-\frac{2}{3}\,{\frac {v_{{g}} \left( {x_{{0}}}^{3/2}-{x}^{3/2} \right) }{
\sqrt {x_{{0}}}v_{{0}}}}+y_{{0}}
\quad .
\label{yxhyperbolic}
\end{equation}
The trajectory
which represents the best fit
for the right side
of \htrg  in the case of an hyperbolic  density profile
is shown in Figure \ref{ngc1265_traj_power1}.
% figure   ngc1265_traj_power1
% end ngc1265_traj_power1

The radius of curvature in the case of
an hyperbolic profile of  density
as  given by
Eq. (\ref{radiuscurvature})
is
\begin{equation}
\curvature=
2\,{\frac { \left( x{v_{{g}}}^{2}+{v_{{0}}}^{2}x_{{0}} \right) ^{3/2}
\sqrt {x}}{{v_{{0}}}^{2}x_{{0}}v_{{g}}}}
\quad ,
\end{equation}
 and the centripetal acceleration is
\begin{align}
\frac{v^2}{\curvature}{\veczan{N}} &= ~ \nonumber \\
\frac
{
 \left(A+ \left( 12\,tv_{{0}}{x_{{0}}}^{2}+8\,{x_{{0}}}^{3} \right) ^{
4/3}+16\,{x_{{0}}}^{2}{y_{{0}}}^{2} \right) {v_{{0}}}^{2}x_{{0}}\sqrt
{4}v_{{g}}
}
{
32\,{x_{{0}}}^{2} \left( {v_{{0}}}^{2}x_{{0}}+1/4\,{\frac {{v_{{g}}}^{
2} \left( 12\,tv_{{0}}{x_{{0}}}^{2}+8\,{x_{{0}}}^{3} \right) ^{2/3}}{x
_{{0}}}} \right) ^{3/2}\sqrt {{\frac { \left( 12\,tv_{{0}}{x_{{0}}}^{2
}+8\,{x_{{0}}}^{3} \right) ^{2/3}}{x_{{0}}}}}
}
{\veczan{N}}
& ~
\quad ,
\end{align}
where
\begin{equation}
A=16\,{t}^{2}{v_{{g}}}^{2}{x_{{0}}}^{2}+32\,tv_{{g}}{x_{{0}}}^{2
}y_{{0}}
\quad .
\end{equation}

\section{The helicoidal trajectory}
\label{sechelicoidal}
The circular helix is
\begin{subequations}
\begin{align}
x= a \,cos (t) \\
y= a \,sin (t) \\
z= b \,t \quad
\end{align}
\end{subequations}
where $a$ is the radius and the pitch is $2\,\pi\, b$,
see \cite{Lipschutz1969}.
The radius of curvature  of the circular helix is
\begin{equation}
\curvature = \frac{a^2 +b^2}{a}
\quad .
\end{equation}
The arc-length  of the helix, $s$,
as a function of time is
\begin{equation}
s= \sqrt{a^2 +b^2} \,t
\quad .
\end{equation}

\begin{figure*}
\begin{center}
\includegraphics[width=7cm]{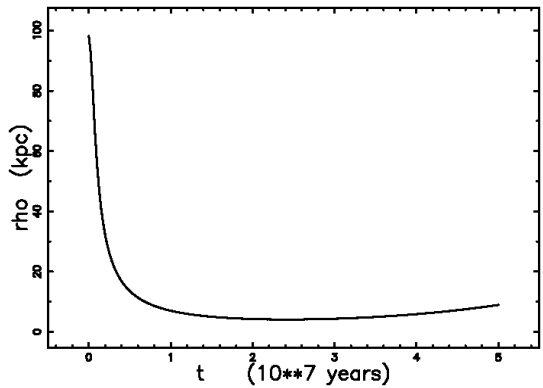}
\end {center}
\caption
{
The radius of curvature of the helicoidal jet
versus time
when  the parameters are
$x_0$=100\,pc,
$y_0$=0\,pc,
and
$v_0=$9620\, km/s.
}
\label{helicoidal_radius}
    \end{figure*}

The helicoidal jet  in the case of constant density is
\begin{subequations}
\label{helicoidaljet}
\begin{align}
x=\sqrt {2\,tv_{{0}}x_{{0}}+{x_{{0}}}^{2}}   \\
y=at\cos \left( \Omega_{{p}}t \right)        \\
z=at\sin \left( \Omega_{{p}}t \right) \quad   ,
\end{align}
\end{subequations}
where the $x-t$  relationship is given by
Eq. (\ref{xtconstant}),
$\Omega_{{p}}$ is the angular velocity, and the
radius of the helix  grows linearly with time
according to a linear relationship given by the parameter $a$.
The  pitch  of the  helicoidal jet is
\begin{equation}
\sqrt {4\,{\frac {\pi \,v_{{0}}x_{{0}}}{\Omega_{{p}}}}+{x_{{0}}}^{2}}
\quad .
\label{pitchhelicoidal}
\end{equation}
The radius of curvature of the helicoidal jet is
\begin{equation}
\curvature = \frac{NH}{DH}
\label{curvaturehelicoidal}
\quad ,
\end{equation}
where
\begin{equation}
NH= \left( 2\,{a}^{2}{t}^{3}{\Omega_{{p}}}^{2}v_{{0}}+{a}^{2}{t}^{2}{
\Omega_{{p}}}^{2}x_{{0}}+2\,{a}^{2}tv_{{0}}+{a}^{2}x_{{0}}+{v_{{0}}}^{
2}x_{{0}} \right) ^{3/2}
\quad ,
\end{equation}
\begin{align}
DH= ~\nonumber \\
a\Big (8\,{a}^{2}{t}^{7}{\Omega_{{p}}}^{6}{v_{{0}}}^{3}+12\,{a}^{2}{t
}^{6}{\Omega_{{p}}}^{6}{v_{{0}}}^{2}x_{{0}}+6\,{a}^{2}{t}^{5}{\Omega_{
{p}}}^{6}v_{{0}}{x_{{0}}}^{2}+{a}^{2}{t}^{4}{\Omega_{{p}}}^{6}{x_{{0}}
}^{3}
\nonumber \\
+32\,{a}^{2}{t}^{5}{\Omega_{{p}}}^{4}{v_{{0}}}^{3}+48\,{a}^{2}{t}
^{4}{\Omega_{{p}}}^{4}{v_{{0}}}^{2}x_{{0}}+4\,{t}^{4}{\Omega_{{p}}}^{4
}{v_{{0}}}^{4}x_{{0}}
+24\,{a}^{2}{t}^{3}{\Omega_{{p}}}^{4}v_{{0}}{x_{{0
}}}^{2}
\nonumber \\
+4\,{t}^{3}{\Omega_{{p}}}^{4}{v_{{0}}}^{3}{x_{{0}}}^{2}+4\,{a}^
{2}{t}^{2}{\Omega_{{p}}}^{4}{x_{{0}}}^{3}+{t}^{2}{\Omega_{{p}}}^{4}{v_
{{0}}}^{2}{x_{{0}}}^{3}+32\,{a}^{2}{t}^{3}{\Omega_{{p}}}^{2}{v_{{0}}}^
{3}
\nonumber \\
+48\,{a}^{2}{t}^{2}{\Omega_{{p}}}^{2}{v_{{0}}}^{2}x_{{0}}+21\,{t}^{
2}{\Omega_{{p}}}^{2}{v_{{0}}}^{4}x_{{0}}+24\,{a}^{2}t{\Omega_{{p}}}^{2
}v_{{0}}{x_{{0}}}^{2}
\nonumber \\
+18\,t{\Omega_{{p}}}^{2}{v_{{0}}}^{3}{x_{{0}}}^{2
}+4\,{a}^{2}{\Omega_{{p}}}^{2}{x_{{0}}}^{3}+4\,{\Omega_{{p}}}^{2}{v_{{0
}}}^{2}{x_{{0}}}^{3}+{v_{{0}}}^{4}x_{{0}}\Big )^{1/2}
\, .
\end {align}
Figure \ref{helicoidal_radius} shows
the radius of curvature of the helicoidal jet in kpc
as a function of time in units of $10^7$ years.
% figure   helicoidal_radius
% end helicoidal_radius

\begin{figure}
 \begin{center}
\includegraphics[width=7cm]{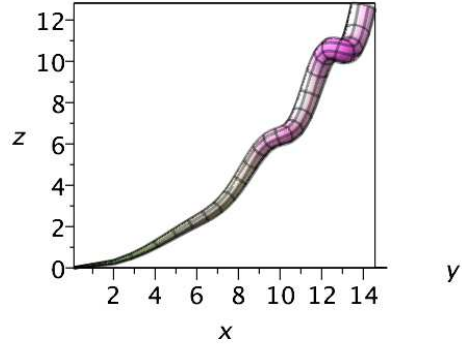}
 \end {center}
\caption
{
 Continuous 3D surface of \htrg:
 the three Eulerian angles
 characterizing the point of view are
   $ \Phi   $= 0  $^{\circ }$,
   $ \Theta $= 90 $^{\circ }$,
 and
   $ \Psi   $= 0  $^{\circ }$.
The physical   parameters are
$x_0$=100 pc,
$y_0$=0  pc,
$v_0=$9620 km/s,
$v_{g}=$124 km/s,
$\Omega_{{p}}=6 \,10^{-8} \,  \pi\, \frac{rad}{years}$
and
$alpha = 5 ^{\circ }$.
The axes are expressed in kpc units.
}
\label{bented3dtube1}
\end{figure}

The bent helicoidal jet  in the case of
constant density is
\begin{subequations}
\begin{align}
x=\sqrt {2\,tv_{{0}}x_{{0}}+{x_{{0}}}^{2}} \\
y=at\cos \left( \Omega_{{p}}t \right)      \\
z=at\sin \left( \Omega_{{p}}t \right) + z_0 +v_g t \quad .
\end{align}
\end{subequations}
A 3D display of the bent helicoidal jet
is shown in Figure \ref{bented3dtube1}, where
the choice of the Euler angles, which define  the observer,
corresponds to the astronomical observations.
%begin figure bented3dtube1
%fine figure bented3dtube1

\begin{figure}
 \begin{center}
\includegraphics[width=7cm]{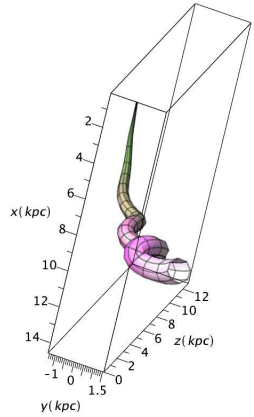}
 \end {center}
\caption
{
 Continuous 3D surface of \htrg:
 the three Eulerian angles
 characterizing the point of view are
   $ \Phi   $= 70 $^{\circ }$,
   $ \Theta $= 20 $^{\circ }$,
 and $ \Psi $= 10 $^{\circ }$.
The physical parameters are the same as in Figure
\ref{bented3dtube1} and
the axes are expressed in kpc units.
}
\label{bented3dtube2}
\end{figure}

Another choice of the Euler angles produces
a different projected surface, see Figure \ref{bented3dtube2}.
%begin figure bented3dtube2
%fine figure bented3dtube2

\begin{figure*}
\begin{center}
\includegraphics[width=7cm]{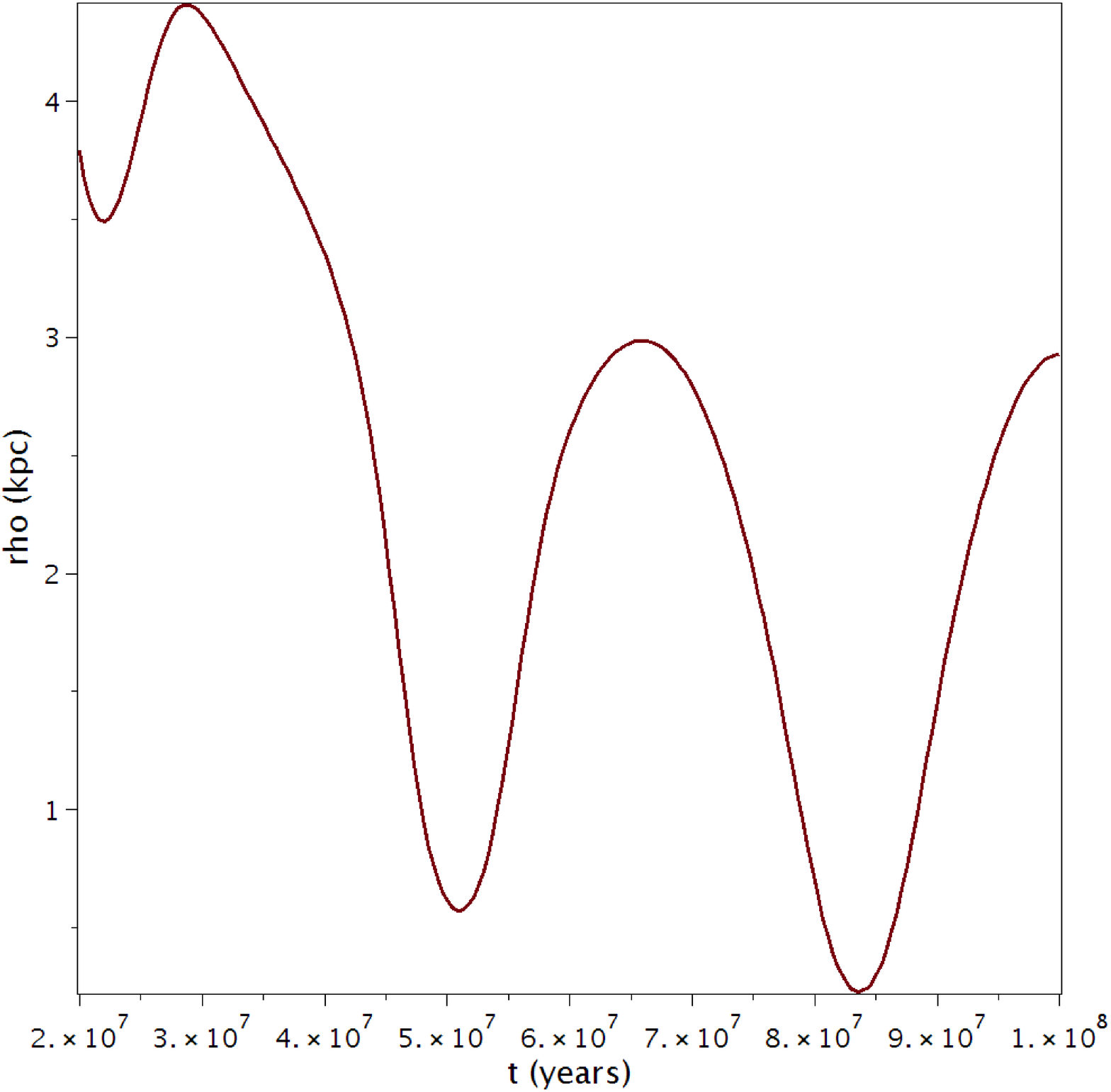}
\end {center}
\caption
{
The radius of curvature of the bent helicoidal jet
versus time.
The physical parameters are the same as Figure
\ref{bented3dtube1}.
}
\label{bentedradius}
    \end{figure*}

The radius  of curvature  of the bent helicoidal jet
has a complicated expression and we only present
its  numerical behavior, see Figure \ref{bentedradius}.
% figure   bentedradius
% end bentedradius

The arc-length  of the bent helicoidal jet is
\begin{equation}
s= \int_0^t \sqrt{
\frac{NS}{DS}\, dt
}
\quad ,
\label{archelicoidalbent}
\end{equation}
where
\begin{eqnarray}
NS=
2\,{a}^{2}{\Omega_{{p}}}^{2}{t}^{3}v_{{0}}+{a}^{2}{\Omega_{{p}}}^{2}{t
}^{2}x_{{0}}+4\,\cos \left( \Omega_{{p}}t \right) a\Omega_{{p}}{t}^{2}
v_{{0}}v_{{g}}+2\,{a}^{2}tv_{{0}}
\nonumber \\
+2\,\cos \left( \Omega_{{p}}t
 \right) a\Omega_{{p}}tv_{{g}}x_{{0}}+{a}^{2}x_{{0}}+4\,\sin \left(
\Omega_{{p}}t \right) atv_{{0}}v_{{g}}
\nonumber \\
+2\,\sin \left( \Omega_{{p}}t
 \right) av_{{g}}x_{{0}}+2\,tv_{{0}}{v_{{g}}}^{2}+{v_{{0}}}^{2}x_{{0}}
+{v_{{g}}}^{2}x_{{0}}
\quad ,
\end{eqnarray}
and
\begin{equation}
DS=
2\,tv_{{0}}+x_{{0}}
\quad .
\end{equation}
The  previous integral  does not  have
an analytical  expression and the  numerical
integration between 0 and $10^8$ years, the other
parameters as  in
Figure~\ref{bentedradius},
gives $s=22.2 \,$pc.

\section{The theory of images}

\label{sectionimage}
\begin{figure}
 \begin{center}
\includegraphics[width=7cm]{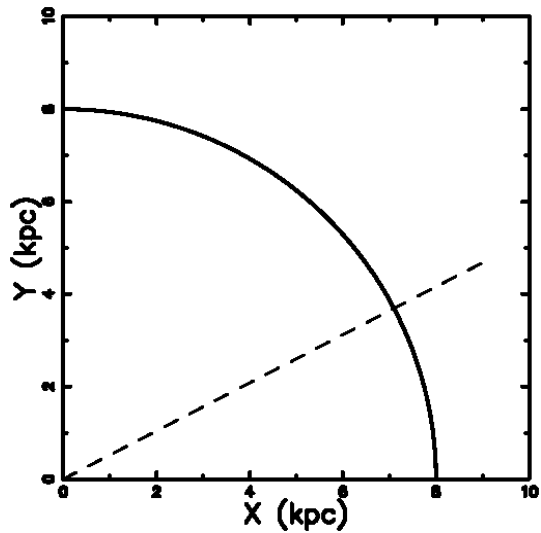}
 \end {center}
\caption {
The source is represented through
a circular section perpendicular to the jet axis.
The observer is situated along the $y$-direction,
one line of sight is indicated
and the angle $\beta$ is clearly indicated.
 }%
 \label{crossview}
 \end{figure}

In  the case of an optically thin layer,
the emissivity is proportional to the number density, $C$,
\begin{equation}
j_{\nu} \zeta =K  C(s)
\quad  ,
\end{equation}
where
$j_{\nu}$ is the emission coefficient
and  $K$ is a  constant.
This can be the case for
synchrotron radiation in the presence of an
isotropic distribution of
electrons with a power law distribution in energy, $N(E)$,
\begin{equation}
N(E)dE = K_s E^{-\gamma} \label{spectrum} \quad  ,
\end{equation}
where $K_s$  and $\gamma$ are  constants, see \cite{Zaninetti2010b}
for more details.
In the case of constant number density
\begin{equation}
j_{\nu} \zeta =K\,C\, s
\quad ,
\end{equation}
where $s$  is the length of the relevant line of sight, this
formula is extremely simple and allows
building the image in the appropriate
geometrical environment.
Two analytical results outline the theoretical framework that
should be verified by the simulation.
The first one analyses
the behavior of the  intensity or brightness
along  a jet when the distance from the origin, $x$,
is fixed.
We assume that the number density $C$ is constant
in a cross section of radius $a$
and then falls to 0, see Figure~\ref{crossview}.
% figure crossview
% end figure crossview

\begin{figure}
 \begin{center}
\includegraphics[width=7cm]{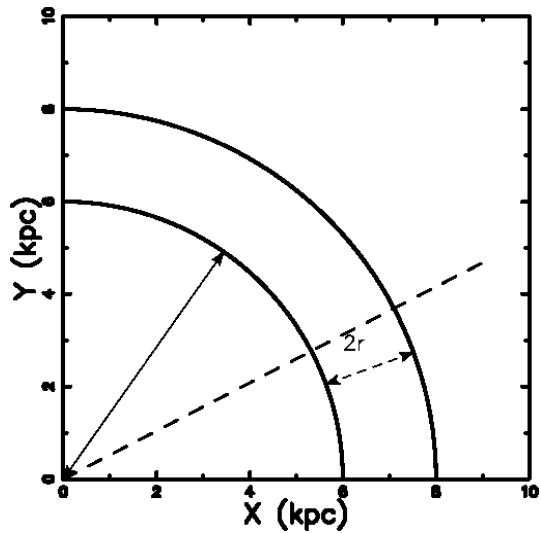}
 \end {center}
\caption {
The section of one-fourth of a toroidal  helix
is represented through
a circle of radius $R-r$ and a
bigger circle of radius $R+r$.
The observer is situated along the $y$-direction at infinity
and  the line of sight of maximum length is indicated.
 }%
 \label{helixview}
 \end{figure}

The length of sight, when the observer is situated
at the infinity of the $y$-axis,
is the locus
parallel to the $y$-axis which crosses the position $x$ in a
Cartesian $x-y$ plane and terminates at the external circle
of radius $a$.
The locus length is
\begin{eqnarray}
l_{ab} = 2 \times ( \sqrt {a^2 -x^2})
\quad ; 0 \leq x < a \quad .
\label{lengthcylinder}
\end{eqnarray}
When the number density $C$ is constant on a cylinder
of radius $a$,
the  intensity or brightness  of radiation is
\begin{eqnarray}
I_{0a} =C \times 2 \times ( \sqrt { a^2 -x^2})
 \quad ; 0 \leq x < a \quad ,
\label{icylinder}
\end{eqnarray}
or
\begin{eqnarray}
I_{0a} =C \times 2 \times a \times \sin (\beta)
 \quad ;-\frac{\pi}{2} \leq \beta \leq \frac{\pi}{2} \quad,
\label{icylindera}
\end{eqnarray}
which  can be called
the `trigonometrical law' for the  intensity or brightness.
The second  analytical  result
can be derived  when
the curved shape of a jet of finite cross section
is  parametrized by a  toroidal helix.
The toroidal helix
has the following parametric equations:
\begin{eqnarray}
x = \cos(\alpha) \cdot (R + r \cdot \cos(\theta)) \nonumber \\
y = \sin(\alpha) \cdot (R + r \cdot \cos(\theta)) \\
z = r \cdot \sin(\theta) + \alpha \frac{ n \, r}{2\pi}
\quad , \nonumber \\
\end{eqnarray}
where $ \theta   \in [0,2\pi)$,
$ \alpha   \in [0,2\pi)$,
$n \, r $ is the distance
along the $z$-axis  after an angle $\alpha=2 \pi$, and $n$  is an integer.
We now analyse the case in which
\begin{equation}
\alpha \ll  \frac {2\pi}{n}
\quad ,
\end{equation}
where the right-hand side  is the value of the angle
after which the toroidal helix  has advanced
by $r$ along the $z$-direction.
Figure~\ref{helixview} shows  a section
in the middle of the toroidal helix $z=0$,
from which is possible to see
that the dotted line presents the
longest line of sight, $l_{max}$,
which starts  from $x=R$
when the observer is at the infinity of the $y$-axis.
% figure helixview
% end figure helixview

The shortest line of sight is $2r$.
The maximum enhancement in the presence
of a constant number density, $e$, is
\begin{equation}
e =\frac {l_{max}} {2r}.
\end{equation}
A simple geometrical demonstration gives
\begin{equation}
e =\frac{ 1}{2} \,{\frac {\sqrt {2\,R+r}}{\sqrt {r}}}
\quad .
\label{fattoree}
\end{equation}
The relationship  between the radius of curvature, $R$,
and the radius of the toroidal helix, $r$,
which   produces an enhancement $e$ in the
intensity or brightness
is
\begin{equation}
\frac{R}{r} = 2\,{e}^{2}-\frac{1}{2}
\quad .
\end{equation}

We now outline how it is  possible to build a radio-image.
The number density is stored  on a 3D   grid ${\mathcal
M(i,j,k)}$ where $i,j$ and $k$ are indices varying from 1 to
$pixels$.
The orientation  of the object is characterized by
the Euler angles $(\Phi, \Theta, \Psi)$
and  therefore  by a
$3 \times 3$  rotation matrix,
$E$, see \cite{Goldstein2002}.
The grid ${\mathcal M(i,j,k)}$ is then rotated according to the chosen
Euler angles.
The intensity map is obtained by summing the points of the
rotated images
along a particular direction,
and the intensity is
\begin{eqnarray}
{\it I}\/(i,j) = \sum_k  \triangle\,s \times  {\mathcal M}(i,j,k)
\\
\quad  \mbox {optically thin layer}
\quad linear~case
\quad,   \nonumber
\label{matrix2d}
\end{eqnarray}
where $\triangle$s is the spatial interval between the various
values of intensity and  the sum is performed over the
interval of existence of the index $k$. In this grid
the little squares that are characterized by the position of
the indexes $i,j$  correspond to a different line of sight.

The effect of the  insertion of a threshold intensity, $ I_{tr} $,
given by the observational techniques, is now analysed.
The threshold intensity can be
parametrized  to  $ I_{max} $,
the maximum  value  of intensity
characterizing the map,
\begin{equation}
I_{tr} =\frac{I_{max}} {factor}
\quad ,
\end{equation}
where the parameter factor is greater than one.

\begin{figure}
 \begin{center}
\includegraphics[width=7cm]{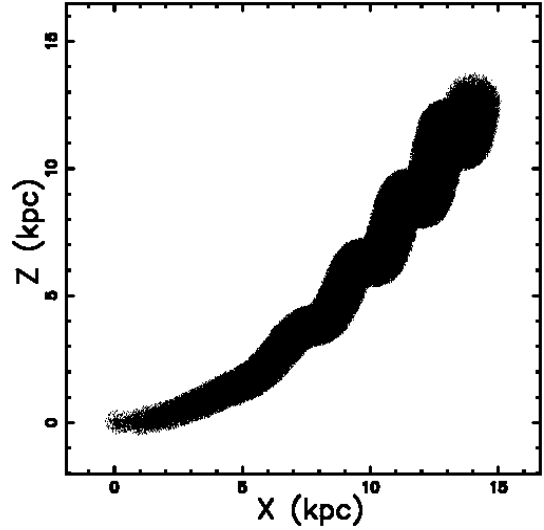}
 \end {center}
\caption {
 Continuous 3D trajectory of \htrg:
 the three Eulerian angles
 characterizing the point of view are
   $ \Phi   $= 0 $^{\circ }$,
   $ \Theta $= 90 $^{\circ }$,
 and $ \Psi $= 0 $^{\circ }$.
The physical parameters are the same as Figure
\ref{bented3dtube1}.
}
\label{proj_ngc1265_molti}
\end{figure}

\begin{figure}
 \begin{center}
\includegraphics[width=7cm]{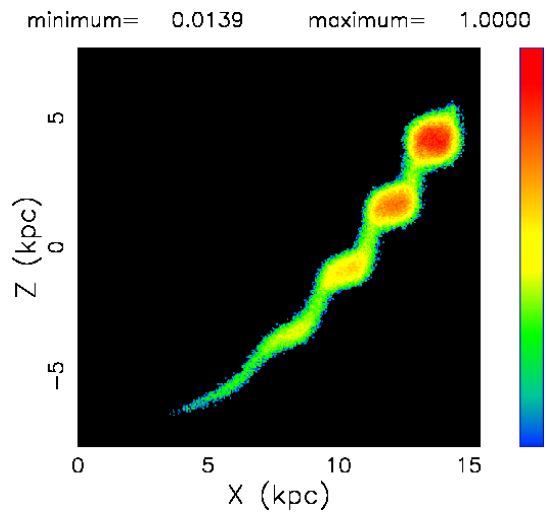}
 \end {center}
\caption {
Theoretical 2D map of the  surface
brightness  of the emission of  \htrg
with basic parameters as in Figure~\ref{proj_ngc1265_molti}.
The integration is performed on a cubic grid
of $1000^3$ pixels and factor = 4.
 }%
 \label{ngc1265_intensity_hole}
 \end{figure}

A map for the emissivity of a jet can be built
by employing the following  algorithm:
\begin{itemize}
\item  A great  number of points, for example, one million,
is  inserted into a bent helicoidal jet with
given characteristic parameters.
\item
This set of points is rotated according to the three Euler angles
which  identify  the observer's point of view.
\item
The 2D matrix  which represents the intensity of the non-thermal emission
is built according to  procedure (\ref{matrix2d}).
\end{itemize}
Figure   \ref{proj_ngc1265_molti}   shows the projected
points.
%begin figure proj_ngc1265_molti
%fine figure proj_ngc1265_molti
The two analytical results, the enhancement  of the intensity
in the central line  and the knot structure
due to the curvature along the line of sight, are clearly
visible  in Figure \ref{ngc1265_intensity_hole}, which shows
the intensity of a non-thermal map  of  \htrg.
% figure ngc1265_intensity_hole
% end figure ngc1265_intensity_hole

A first  comparison  of the previous figure can be done
with Figure 1 in  \cite{Aloy2003}  where a 3D relativistic
hydrodynamic simulation for the  precessing beam
was carried out.
A second  comparison  can be done  with the pseudo-synchrotron
intensities  visible in the figures of \cite{Hardee2003,Hardee2005}
where the knots in the radio-jets   were  simulated in the framework
of  helical relativistic instabilities.
A third comparison can be done with Figure 3 in \cite{Laing2014}
where the  observed and model total-intensity images
are reported for 15 radio-galaxies without bending effects.

\section{Conclusions}

{\bf Classical turbulence:}
Turbulent jets  are usually modeled by a temporal evolution
with density equal to that of the surrounding medium.
Here, in order to cover the astrophysical applications,
we considered an hyperbolic profile of density,
see the solution (\ref{xthyperbolic}) and an asymptotic solution
(\ref{xthyperbolicasympt}).
The case of a density which follows an inverse power law
is limited to the derivation of the velocity, see
Eq. (\ref{velocityhyper}).
This inverse power law case allows matching a wide variety
of astrophysical situations, as an example, when $\delta=2$, the
velocity does not decreases with distance.

This is obvious from the conservation equation,
$\rho vA$ = constant.
If $A$ goes as $r^2$ and $\rho$
goes with the inverse of the square of distance in
a conical jet, $v$ is constant.

{\bf Relativistic turbulence:}
The conservation of the relativistic momentum flux  for turbulent
jets is here analysed in three cases.
The first case is that with a surrounding medium having constant density,
where the analytical result   is limited to
a series expansion for the solution, see Eq. (\ref{rtseries}).
The second case is that of an hyperbolic density decrease for the
surrounding medium, for which  we derived an analytical solution
see Eq. (\ref{xtrelativistic}) and an  asymptotic solution, see Eq.
(\ref{xtrelativistichyperbolic}).
The third case is that where the surrounding density
decreases with a power law behavior:  the analytical result
is limited to the velocity--distance relationship,
see Eq. (\ref{betadistance}).

{\bf The curvature of the jet:}
The composition of the velocity is discussed in the light
of the radius of curvature and the standard  mathematical
definition is given, see Eq. (\ref{rhogeometrical}).
The astronomical  counterpart  is the radius of the circle
of curvature which is shown in Figure~\ref{ngc1265_circle}
for \htrg.
Two analytical solutions are given for the curved trajectory,
see Eqs (\ref{xft}) and (\ref{yft}),
in the case of constant density
and  Eq. (\ref{yxhyperbolic}) for an hyperbolic decrease in density.
A careful analysis  of Figures
\ref{ngc1265_traj}
and
\ref{ngc1265_traj_power1}   does not show an accurate coincidence
between the predictions of the model and the digitized trajectory.
In this case, its $\chi^2$ can be lowered by introducing other effects
such as photon losses due to synchrotron radiation
 or another density profile.

{\bf Helicoidal bent jet:}
The precession  is here modeled by an helicoidal jet, see
Eq. \ref{helicoidaljet}, which has a pitch
and radius of curvature given by
Eqs (\ref{pitchhelicoidal}) and (\ref{curvaturehelicoidal}).
The composition with the  velocity of the host galaxy
allows modelling the
a helicoidal jet in which the  radius of curvature
can be visualized only numerically,
see  Figure \ref{bentedradius}.
The arc-length  has a complicated expression, which  is given
by  Eq. (\ref{archelicoidalbent}).

{\bf Theoretical radio maps:}
The radio image  of an extragalactic radio source
is built  adopting: (i) a uniform number density of
synchrotron emitters over the entire jet, (ii) the thin layer approximation.
The two theoretical effects  of central brightening, see
Eq. (\ref{icylindera}), and intensity brightening  due
to the curvature  of the emitting region, see Eq. (\ref{fattoree}),
are both visible in the simulation of  \htrg,
see Figure \ref{ngc1265_intensity_hole}.

New analytical results can be obtained expressing the
helicoidal trajectory in cylindrical coordinates or
analysing the `valley on the top' effect which is due to the
velocity profile in the direction perpendicular to the motion,
see \cite{Zaninetti2009d}.
The previously cited papers solve some of the
problems connected with the radio-jets
but leave other problems open,
and  Table~\ref{problems} reports their status.
\begin{table}
 \caption {Synoptic table of the assumptions of two papers and here }
 \label{problems}
 \[
 \begin{array}{lccc}
 \hline
 \hline
 \noalign{\smallskip}
Problem  &  Laing \& Bridle \,2002
& Hardee\, et\,al.  2005
&
this~paper      \\
 \noalign{\smallskip}
 \hline
 \noalign{\smallskip}
distance~  along~ the~ jet & numerical & not &  analytical \\
velocity~  along~ the~ jet & numerical & not &  analytical \\
knots                      & not       & instabilities & image~theory\\
\noalign{\smallskip}
\noalign{\smallskip}
  \hline
 \end{array}
 \]
 \end {table}

\appendix

\section {The circle of curvature}
\label{appendice}
\setcounter{equation}{0}
\renewcommand{\theequation}{\thesection.\arabic{equation}}

Once three points are selected on a curve,
$(x_1,y_1) , (x_2,y_2), (x_3,y_3)$,
the circle of curvature has radius
\begin{equation}
\circleradius = \frac{NR}{DR}
\label{circleradius}
\quad ,
\end{equation}
where
\begin{align}
NR=
\Bigg \{ \left( {x_{{2}}}^{2}-2\,x_{{2}}x_{{3}}+{x_{{3}}}^{2}+{y_{{2}}}
^{2}-2\,y_{{2}}y_{{3}}+{y_{{3}}}^{2} \right) \times &
\nonumber \\
\left( {x_{{1}}}^{2}-2\,
x_{{1}}x_{{3}}+{x_{{3}}}^{2}+{y_{{1}}}^{2}-2 y_{{1}}y_{{3}}+{y_{{3}}}
^{2} \right) \times & \nonumber \\
 \left( {x_{{1}}}^{2}-2\,x_{{1}}x_{{2}}+{x_{{2}}}^{2}+{y_
{{1}}}^{2}-2\,y_{{1}}y_{{2}}+{y_{{2}}}^{2} \right) \Bigg \}^{(1/2)} &
\quad ,
\nonumber
\end{align}
\begin{equation}
DR=2\,x_{{1}}y_{{2}}-2\,x_{{1}}y_{{3}}-2\,x_{{2}}y_{{1}}+2\,x_{{2}}y_{{3}
}+2\,x_{{3}}y_{{1}}-2\,x_{{3}}y_{{2}}
\nonumber
\quad .
\end{equation}
The two coordinates of the centre of the circle,
($x_c,y_c$),
are
\begin{equation}
x_c = \frac{NXC}{DXC}
\quad ,
\label{circlexc}
\end{equation}
where
\begin{align}
NXC=
{x_{{1}}}^{2}y_{{2}}-{x_{{1}}}^{2}y_{{3}}-{x_{{2}}}^{2}y_{{1}}+{x_{{2}
}}^{2}y_{{3}}+{x_{{3}}}^{2}y_{{1}}-{x_{{3}}}^{2}y_{{2}}+{y_{{1}}}^{2}y
_{{2}}
\nonumber \\
-{y_{{1}}}^{2}y_{{3}}-y_{{1}}{y_{{2}}}^{2}+y_{{1}}{y_{{3}}}^{2}+
{y_{{2}}}^{2}y_{{3}}-y_{{2}}{y_{{3}}}^{2}
\quad ,
\nonumber
\end{align}
\begin{equation}
DXC=
2\,x_{{1}}y_{{2}}-2\,x_{{1}}y_{{3}}-2\,x_{{2}}y_{{1}}+2\,x_{{2}}y_{{3}
}+2\,x_{{3}}y_{{1}}-2\,x_{{3}}y_{{2}}
\quad ,
\nonumber
\end{equation}
and
\begin{equation}
y_c = \frac{NYC}{DYC}
\quad ,
\label{circleyc}
\end{equation}
where
\begin{align}
NYC=
-{x_{{1}}}^{2}x_{{2}}+{x_{{1}}}^{2}x_{{3}}+x_{{1}}{x_{{2}}}^{2}-x_{{1}
}{x_{{3}}}^{2}+x_{{1}}{y_{{2}}}^{2}-x_{{1}}{y_{{3}}}^{2}
\nonumber \\
-{x_{{2}}}^{2}
x_{{3}}+x_{{2}}{x_{{3}}}^{2}-x_{{2}}{y_{{1}}}^{2}+x_{{2}}{y_{{3}}}^{2}
+x_{{3}}{y_{{1}}}^{2}-x_{{3}}{y_{{2}}}^{2}
\quad ,
\nonumber
\end{align}
\begin{equation}
DYC=
2\,x_{{1}}y_{{2}}-2\,x_{{1}}y_{{3}}-2\,x_{{2}}y_{{1}}+2\,x_{{2}}y_{{3}
}+2\,x_{{3}}y_{{1}}-2\,x_{{3}}y_{{2}}
\quad .
\nonumber
\end{equation}

%\bibliography{biblio}

\end{document}